\newif\iffigs\figstrue
\documentclass[12pt]{article}
   \input{epsf}
\else
\fi

\newcommand{\sect}[1]{\setcounter{equation}{0}\section{#1}}

\usepackage{amsmath}
\usepackage{amscd}
\usepackage{array}
\usepackage{latexsym,amssymb}

\begin{document}


\newcommand{\nn}{\nonumber}
\newcommand{\mm}{\hfill\cr}
\newcommand{\noi}{\noindent}
\def\di{\displaystyle}
\def\sx{\left}
\def\dx{\right}
\def\to{\rightarrow}
\def\ul{\underline}

\newcommand{\bm}[1]{\mbox{\boldmath $#1$}}

\newcommand{\be}{\begin{equation}}
\newcommand{\ee}{\end{equation}}
\newcommand{\bea}{\begin{eqnarray}}
\newcommand{\eea}{\end{eqnarray}}
\newcommand{\ov}{\overline}
\newcommand{\ba}{\begin{eqnarray}}
\newcommand{\ea}{\end{eqnarray}}

\def\sk{\vskip .4cm}

\def\a{\alpha}
\def\ap{\alpha'}
\def\b{\beta}
\def\c{\chi}
\def\cb{\ol {\chi}}
\def\d{\delta}
\def\e{\epsilon}
\def\f{\varphi}
\def\g{\gamma}
\def\k{\kappa}
\def\l{\lambda}
\def\m{\mu}
\def\n{\nu}
\def\o{\omega}
\def\p{\psi}
\def\r{\rho}
\def\s{\sigma}
\def\t{\tau}
\def\th{\theta}
\def\ve{\varepsilon}
\def\z{\zeta}

\def\D{\Delta}
\def\G{\Gamma}
\def\L{\Lambda}
\def\O{\Omega}
\def\S{\Sigma}
\def\Th{\Theta}


\newcommand{\rtr}{\mathrm{tr}}
\newcommand{\rU}{\mathrm{U}}
\newcommand{\rUSp}{\mathrm{USp}}
\newcommand{\rSU}{\mathrm{SU}}
\newcommand{\rE}{\mathrm{E}}
\newcommand{\rSO}{\mathrm{SO}}
\newcommand{\rSL}{\mathrm{SL}}
\newcommand{\cV}{\mathcal{V}}
\newcommand{\rSp}{\mathrm{Sp}}
\newcommand{\rF}{\mathrm{F}}
\newcommand{\rGL}{\mathrm{GL}}
\newcommand{\rG}{\mathrm{G}}
\newcommand{\rK}{\mathrm{K}}


\def\cA{\mathcal{A}}
\def\cB{\mathcal{B}}
\def\cC{\mathcal{C}}
\def\cD{\mathcal{D}}
\def\cE{\mathcal{E}}
\def\cF{\mathcal{F}}
\def\cG{\mathcal{G}}
\def\cH{\mathcal{H}}
\def\cI{\mathcal{I}}
\def\cJ{\mathcal{J}}
\def\cK{\mathcal{K}}
\def\cL{\mathcal{L}}
\def\cM{\mathcal{M}}
\def\cN{\mathcal{N}}
\def\cO{\mathcal{O}}
\def\cP{\mathcal{P}}
\def\cQ{\mathcal{Q}}
\def\cR{\mathcal{R}}
\def\cS{\mathcal{S}}
\def\U{\mathcal{U}}
\def\cV{\mathcal{V}}
\def\cW{\mathcal{W}}

\def\lag{{\mathcal{L}}}


\newcommand{\fgl}{\mathfrak{gl}}
\newcommand{\fu}{\mathfrak{u}}
\newcommand{\fsl}{\mathfrak{sl}}
\newcommand{\fsp}{\mathfrak{sp}}
\newcommand{\fusp}{\mathfrak{usp}}
\newcommand{\fsu}{\mathfrak{su}}
\newcommand{\fp}{\mathfrak{p}}
\newcommand{\fso}{\mathfrak{so}}
\newcommand{\fl}{\mathfrak{l}}
\newcommand{\fg}{\mathfrak{g}}
\newcommand{\fr}{\mathfrak{r}}
\newcommand{\fe}{\mathfrak{e}}
\newcommand{\ft}{\mathfrak{t}}


\def\tI{{\Lambda}}
\def\tJ{{\Sigma}}
\def\tK{{\Gamma}}
\def\tL{{\Delta}}
\def\tM{{\tilde M}}
\def\tN{{\tilde N}}
\def\dt{{\tilde d}}
\def\Dt{{\tilde D}}



\def\real{{\rm Re}\hskip 1pt}
\def\imag{{\rm Im}\hskip 1pt}
\def\cc{{\rm c.c.}}
\def\re{{\rm Re}\mathcal{N}}
\def\im{{\rm Im}\mathcal{N}}
\def\ii{\mathrm{i}}
\def\ib{{\ol {\imath}}}
\def\j{\jmath}
\def\jb{{\ol {\jmath}}}
\def\kb{{\ol  k}}
\def\lb{{\ol  \ell}}
\def\mb{{\ol  m}}
\def\nb{{\ol {n}}}
\def\rb{{\ol {r}}}
\def\sb{{\ol {s}}}


\def\trace{{\rm Tr}\hskip 1pt}
\def\diag{{\rm diag}}
\def\notin{\hbox{{$\in$}\kern-.51em\hbox{/}}}

\def\inbar{\vrule height1.5ex width.4pt depth0pt}
\def\IB{\relax{\rm I\kern-.18em B}}
\def\IC{\relax\,\hbox{$\inbar\kern-.3em{\rm C}$}}
\def\ID{\relax{\rm I\kern-.18em D}}
\def\IE{\relax{\rm I\kern-.18em E}}
\def\IF{\relax{\rm I\kern-.18em F}}
\def\IG{\relax\,\hbox{$\inbar\kern-.3em{\rm G}$}}
\def\IH{\relax{\rm I\kern-.18em H}}
\def\II{\relax{\rm I\kern-.17em I}}
\def\IK{\relax{\rm I\kern-.18em K}}
\def\IL{\relax{\rm I\kern-.18em L}}
\def\IN{\relax{\rm I\kern-.18em N}}
\def\IP{\relax{\rm I\kern-.18em P}}
\def\IQ{\relax\,\hbox{$\inbar\kern-.3em{\rm Q}$}}
\def\IR{\relax{\rm I\kern-.18em R}}
\def\IU{\relax\,\hbox{$\inbar\kern-.3em{\rm U}$}}
\def\ZZ{\relax\ifmmode\mathchoice{\hbox{\cmss Z\kern-.4em Z}}{\hbox{\cmss Z\kern-.4em Z}}{\lower.9pt\hbox{\cmsss Z\kern-.4em Z}} {\lower1.2pt\hbox{\cmsss Z\kern-.4em Z}}\else{\cmss Z\kern-.4em Z}\fi}
\def\IGam{\relax{{\rm I}\kern-.18em \Gamma}}

\newcommand{\iden}{\mbox{{1}\hspace{-.11cm}{l}}}
\def\bfnull{\relax{\rm O \kern-.635em 0}}


\def\de{{\rm d}}
\def\der{\partial}
\def\bos{{\rm bos}}
\def\na{\nabla}
\def\ol{\overline}
\def\ot{\otimes}
\def\imez{\frac{{\rm i}}{2}}
\def\mez{\frac{1}{2}}
\def\qu{\frac{1}{4}}
\def\we{\wedge}
\def\square{{\,\lower0.9pt\vbox{\hrule \hbox{\vrule height 0.2 cm \hskip 0.2 cm \vrule height 0.2 cm}\hrule}\,}}

\def\twomat#1#2#3#4{\left(\begin{array}{cc} \end{array} \right)}
\def\twovec#1#2{\left(\begin{array}{c} {#1}\\ {#2}\\ \end{array} \right)}


\def\C{\mathbb{C}}
\def\Z{\mathbb{Z}}
\def\Hb{\mathbb{H}}
\def\Eb{{\bf E}}
\def\Rb{{\bf R}}
\def\Eb{{\bf E}}
\def\gb{{\bf g}}

\begin{titlepage}
\rightline{DISTA/PHYS-014/09}
\rightline{April 2009} \vskip 2em
\begin{center}{\bf TRIALITY INVARIANCE IN THE N=2 SUPERSTRING}
\\[3em]
{\bf Leonardo Castellani}, {\bf Pietro Antonio Grassi}, {\bf and Luca Sommovigo}\\ [2em]
{\sl Dipartimento di Scienze e Tecnologie Avanzate and
\\ INFN Gruppo collegato di Alessandria,\\Universit\`a del Piemonte Orientale,\\ Via Teresa Michel 11, 15121
Alessandria, Italy}\\ [2.5em]
\end{center}

\begin{abstract}
We prove the discrete triality invariance of
the $N=2$ NSR superstring moving in a $D=2+2$ target space.
We find that triality holds also in the Siegel-Berkovits
formulation of the selfdual superstring.
A supersymmetric generalization of Cayley's
hyperdeterminant, based on a quartic invariant of the
$SL(2|1)^3$ superalgebra, is presented.

\end{abstract}

\vskip 10cm \noi \hrule \vskip.2cm \noi {\small
leonardo.castellani, pietro.grassi, luca.sommovigo@mfn.unipmn.it }

\end{titlepage}

\newpage
\setcounter{page}{1}

\sect{Introduction}

Cayley's hyperdeterminant \cite{Cayley}, the generalization to cubic $2 \times 2 \times 2$
matrices of the usual determinant of square $2 \times 2$ matrices, was recently recognized
\cite{Duff0} to be at the basis
of fascinating connections between black hole entropy in string theory and the
quantum entanglement of qubits and qutrits in quantum information theory (see \cite{PR2008}
and references therein).

The hyperdeterminant was also used in \cite{Duff1}  to rewrite
the Nambu-Goto Lagrangian for a $D=4$ target space with signature (2,2) in a way that makes manifest a hitherto hidden discrete symmetry. The eight variables given by the world-sheet derivatives of the string coordinate functions $\partial_\alpha X^\mu$  are rearranged in a $2 \times 2 \times 2$ hypermatrix
$X_{AA'A''}$, whose hyperdeterminant square root is shown to coincide with the Nambu-Goto action.
The hyperdeterminant being invariant under interchange of the indices {\small $A,A',A''$},
the triality invariance of the Nambu-Goto Lagrangian becomes explicit. Moreover the
hyperdeterminant encodes in a symmetric way also the $[SL(2,R)]^3$ symmetry of the action, where
the $SL(2,R)$ acting on the index {\small $A$} and the $SL(2,R)$ acting on the index {\small $A'$}
are the $O(2,2)$ spacetime symmetry, and the  $SL(2,R)$ acting on {\small $A''$}
is the world-sheet symmetry.

In \cite{NR2007} the Green-Schwarz $\sigma$-model for the $N=2$ superstring in
$D=2+2$ target space was  re-expressed in terms of an hyperdeterminant, once the zweibein is eliminated via its
(non-algebraic) field equation. The issue of quantum equivalence of the resulting action with the
original  GS $N=2$ superstring, or with the NSR $N=2$ superstring, is still not completely settled.

In this Letter we make manifest a discrete triality invariance of the NSR $N=2$ superstring moving
in a $D=2+2$ target space, without direct recourse to Cayley's hyperdeterminant, but rearranging
the fields in a way suggested by the hyperdeterminant. This triality could
well be the origin of the triality observed in \cite{OV} between the worldsheet moduli, the complex moduli
of the target, and the metric moduli of the target.

Moreover, considering the Siegel-Berkovits action
for the selfdual superstring \cite{Siegel1,Siegel2,Berkovits},
we find that triality holds also in its matter part.

It is natural to ask whether the NSR N=2 superstring in $D=2+2$ target space
could be expressed in terms of a supersymmetric
generalization of the hyperdeterminant. We present such a
generalization, based on a quartic invariant of the
$SL(2|1)^3$ superalgebra.

\sect{The $N=2$ superstring action}

The $N=2$ NSR superstring action \cite{N2superstring} in a flat
target space of signature (2,2) \footnote{The $D=2+2$ critical
dimension for the $N=2$ superstring was first found in
\cite{daddalizzi}} and in the conformal gauge is given by :
\begin{equation}
S_{N=2}=-\frac{1}{ 2\pi} \int d^2 \sigma ( \partial_\alpha X^\mu
\partial^\alpha X^\nu + \partial_\alpha Y^\mu
\partial^\alpha Y^\nu - i \ol \psi^\mu_i \gamma^\alpha
\partial_\alpha \psi^{\nu}_i)\eta_{\mu\nu} \label{N2actionold}
\end{equation}

\noindent where we have used the notations of \cite{GSW}:
$\eta_{\mu\nu}= (1,-1)$ is the two-dimensional Minkowski metric,
$\mu=0,1$, $i=1,2$ and the $\gamma^\alpha$ are the two-dimensional
Dirac matrices
\begin{equation}
\gamma^0 =
\begin{pmatrix}
0 & -i \\
i & 0 \\
\end{pmatrix},
\qquad \gamma^1 =
\begin{pmatrix}
0 & i \\
i & 0 \\
\end{pmatrix}
\label{rhodefinition}
\end{equation}

\noindent The fermions $\psi^\mu_i$ are two-dimensional Majorana
fermions, i.e.:
 \begin{equation}
 \ol \psi^\mu_i \equiv
(\psi^\mu_i)^\dagger \gamma^0= (\psi^\mu_i)^T \gamma^0
 \end{equation}

\noi $\gamma^0$ being the charge conjugation matrix (see the
Appendix A for conventions).
 The action (\ref{N2actionold}) is invariant under the supersymmetry variations:
\bea
& & \delta X =  \ol \epsilon_i \psi_i \\ & &  \delta Y =
\epsilon_{ij} \ol \epsilon_i \psi_j\\ & & \delta \psi_i = -i
\gamma^\alpha
\partial_\alpha X \epsilon_i + i \epsilon_{ij}\gamma^\alpha
\partial_\alpha Y \epsilon_j \label{susyvarold} \eea

We can rearrange the
bosonic and fermionic degrees of freedom (respectively
$X^\m$, $Y^\m$ and $\psi_i{}^\m$) in the  $2 \times 2$ matrices:

\begin{equation}
X_{AA'} \equiv \frac{1}{\sqrt{2}}
\begin{pmatrix}
- X^0 + X^1 & Y^0 - Y^1 \\
- Y^0 - Y^1 & - X^0 - X^1
\end{pmatrix}
\label{xaadef}
\end{equation}
and
\begin{equation}
\psi_{AA'} \equiv \frac{1}{\sqrt{2}}
\begin{pmatrix}
- \psi_1{}^0 + \psi_1{}^1 & & \psi_2{}^0 - \psi_2{}^1 \\
- \psi_2{}^0 - \psi_2{}^1 & & - \psi_1{}^0 - \psi_1{}^1
\end{pmatrix}
\label{psiaadef}
\end{equation}

\noindent Using these notations, the Lagrangian in (\ref{N2actionold}) can be recast in the form:

\begin{equation}
\cL_{N=2} = \sx( X_{AA'A''} X_{BB'B''} - i \ol \psi_{AA'} \g_{A''} \partial_{B''} \psi_{BB'} \dx) \sx( \eta^{A''B''} \e^{AB} \e^{A'B'} \dx)
\label{susylag}
\end{equation}

\noindent with $X_{AA'A''} \equiv \partial_{A''} X_{AA'}$.

\noindent The supersymmetry variations (\ref{susyvarold}) become:
\begin{equation}
\begin{split}
\d X_{AA'} &= \ol \e_i \rho_i{}_{A'}{}^{B'} \psi_{AB'} \\
\d \psi_{AA'} &= -i \der_{A''} X_{AB'} {\rho^t}_i{}^{B'}{}_{A'} \g^{A''} \e_i
\end{split}
\label{susyvars}
\end{equation}
where the  $2 \times 2$ matrices $\rho_i$ are:
\begin{equation}
\rho_1 =
\begin{pmatrix}
1 & 0 \\
0 & 1 \\
\end{pmatrix},
\qquad \rho_2 =
\begin{pmatrix}
0 & -1 \\
1 & 0 \\
\end{pmatrix}
\label{rhodefinition1}
\end{equation}

\sect{Triality invariance}

Consider now the world-sheet metric:
\begin{equation}
G''_{A''B''}
\equiv \partial_{A''} X^\mu \partial_{B''} X^\nu \eta_{\mu\nu}
  + \partial_{A''} Y^\mu \partial_{B''} Y^\nu \eta_{\mu\nu} =
   X_{AA'A''} X_{BB'B''} \e^{AB} \e^{A'B'}
\end{equation}

\noindent In terms of $ G''_{A''B''}$ the bosonic part of the Lagrangian
(\ref{susylag}) is given by:
\begin{equation}
{\cal L}_{b}=\eta^{A''B''} G''_{A''B''} \label{bosonicL}
\end{equation}
As shown by Duff in \cite{Duff1}, the Nambu-Goto Lagrangian
\begin{equation}
{\cal L}_{NG}= \sqrt{-\det ( G''_{A''B''})}
\end{equation}
is invariant under the discrete triality transformations
interchanging the three indices of $X_{AA'A''}$. In fact the usual
determinant of the world-sheet metric $G''$ can be reexpressed as
(minus) the Cayley's hyperdeterminant of the cubic matrix
$X_{AA'A''}$, which is explicitly triality invariant \cite{Duff1}:
\begin{equation}
  {\rm Det}  X \equiv - \frac{1}{ 2} \e^{AB} \e^{A'B'} \e^{CD} \e^{C'D'}\e^{A''D''} \e^{B''C''}
  X_{AA'A''} X_{BB'B''} X_{CC'C''} X_{DD'D''}
  \end{equation}

We prove now that also the bosonic Lagrangian ${\cal L}_b$ in
(\ref{bosonicL}) is invariant under triality, up to total divergence terms.
To show this, we need the metrics $G$ and $G'$ defined by:
\bea
& & G_{AB}  \equiv  X_{AA'A''} X_{BB'B''} \e^{A'B'} \e^{A''B''} \\
& & G'_{A'B'}  \equiv  X_{AA'A''} X_{BB'B''} \e^{A''B''} \e^{AB}
\eea

\noindent With these metrics, we can write the ``triality symmetrized" bosonic Lagrangian, explicitly invariant under triality:
\begin{equation}
\begin{split}
\cL_b &= \sx( \eta^{AB} G_{AB} + \eta^{A'B'} G'_{A'B'} + \eta^{A''B''} G''_{A''B''} \dx) = \\
&= \sx( g^{AA'\, , \, BB'} \eta^{A''B''} + B^{AA'\, , \, BB'} \e^{A''B''} \dx) X_{AA'A''} X_{BB'B''}
\end{split}
\label{LCG}
\end{equation}
where
\begin{equation}
g^{AA'\, , \, BB'} = \e^{AB} \e^{A'B'} = g^{BB'\, , \, AA'}
\label{gdef}
\end{equation}
plays the role of a 4-dimensional flat metric, and
\begin{equation}
B^{AA'\, , \, BB'} = \sx( \e^{AB} \eta^{A'B'} + \e^{A'B'} \eta^{AB} \dx) = - B ^{BB'\, , \, AA'}
\label{Bdef}
\end{equation}
plays the role of a 4-dimensional constant $B$--field.

\noindent The triality-invariant Lagrangian  (\ref{LCG}) differs
from ${\cal L}_b$ in
(\ref{bosonicL}) by the $B$--term: this term is a total divergence, since
it is equal to
\begin{equation}
B^{AA'\, , \, BB'} \e^{A''B''}  \partial_{A''} X_{AA'}
\partial_{B''} X_{BB'}=  \partial_{A''} (B^{AA'\, , \, BB'} \e^{A''B''} X_{AA'}
\partial_{B''} X_{BB'})
\end{equation}

This result can be generalized to the supersymmetric case. The Lagrangian

\begin{equation}
\cL_{N=2} = \sx( g^{AA'\, , \, BB'} \eta^{A''B''} + B^{AA'\, , \, BB'} \e^{A''B''} \dx) \sx(X_{AA'A''} X_{BB'B''} - i
\ol \psi_{AA'} \g_{A''} \partial_{B''} \psi_{BB'} \dx)
\label{susyL}
\end{equation}
is explicitly invariant under (\ref{susyvars}) and triality transformations, and differs from the original
$N=2$ superstring Lagrangian in (\ref{susylag}) only by the $B$-terms. Again these terms are a total
divergence. This has already been proven for the $BXX$ term; to show that also the $B \psi \psi$ term is a total divergence we just have to use the antisymmetry of $B$
and the equality:
\begin{equation}
\ol \psi_{AA'} \g_{A''} \partial_{B''} \psi_{BB'}=  - \partial_{B''} \ol \psi_{BB'} \g_{A''} \psi_{AA'}
\end{equation}

\noi due to $\psi$ being a $D=2$ Majorana fermion.

\section{Siegel-Berkovits formulation}

As shown by Siegel \cite{Siegel1,Siegel2} one can describe self-dual
super-Yang-Mills in superspace by extending the bosonic coordinates $X_{AA'}$ to
$(X_{AA'}, \Theta_{A j})$ where $j=1, \dots, {\cal N}$. (we do not include the antichiral coordinates
as in \cite{Berkovits}). It is convenient to cast the $(X_{AA'},\Theta_{A j})$ into a supercoordinate
$Y_{A J} = (X_{AA'},\Theta_{A j})$ with {\small $J=(A',j)$}, which is a vector representation of the supergroup
$OSp({\cal N}|2)$. In order to implement the triality we choose the real form $OSp(2,2|2)$ which has the subgroups $SO(2,2)\times Sp(2) \sim SL(2,R) \times SL(2,R) \times SL(2,R)$. Therefore, the supercoordinates are labelled by $(X_{A A_1}, \Theta_{A A_2 A_3})$ where the $SO(2,2)$ acts on the {\small $A_2$} and {\small $A_3$} indices, $Sp(2)$ acts on {\small $A_1$} and the supersymmetry generators
$Q_{A_1, A_2 A_3}$ act as follows
\begin{equation}\label{SBA}
Q_{A_1, A_2 A_3} X_{B B_1} = \e_{A_1 B_1} \Theta_{B A_2 A_3}\,, ~~~~~
Q_{A_1, A_2 A_3} \Theta_{B B_2 B_3 }= \e_{A_2 B_2}  \e_{A_3 B_3} X_{B A_1}\,.
\end{equation}
Notice that there are effectively four $SL(2,R)$ groups. Let us denote them by
$SL_0(2,R) \times SL_1(2,R) \times SL_2(2,R) \times SL_3(2,R)$. In addition, we
add the $SL(2,R)$ of the worldsheet and we denote it by $SL_w(2,R)$.
We have denoted by {\small $A_0,A_1,A_2,A_3,A_w$} the indices for each of them. 
Thus, for example, the bosonic coordinates $X_{A_0 A_1 A_w}$ transform under 
$SL_0(2,R) \times SL_1(2,R) \times SL_w(2,R)$.

In the formulation of \cite{Berkovits}, the matter part of the action reads
\begin{eqnarray}\label{SBB}
S_m & =& \int d^2z \, \Big( \partial Y_{A J} \bar \partial Y^{A J}
\Big) \\
&=&  \!\! \!\!\!\!\int d^2x\,  \Big( \eta^{A_w B_w} \e^{A_0 B_0} \e^{A_1 B_1}
X_{A_0 A_1 A_w} X_{B_0 B_1 B_w}  \nonumber \\
&+&
\eta^{A_w B_w} \e^{A_0 B_0} \e^{A_2 B_2}  \e^{A_3 B_3}
\Theta_{A_0 A_2 A_3 A_w} \Theta_{B_0 B_2 B_3 B_w}  \Big)\, \nonumber
\end{eqnarray}
\noi where $\Theta_{A_0 A_2 A_3 A_w} \equiv \partial_{A_w} \Theta_{A_0 A_2 A_3 }$.

 The contraction of indices is performed with the invariant tensors of
$SL_0(2,R) \times SL_1(2,R) \times SL_2(2,R) \times SL_3(2,R)$
(except for the worldsheet indices, contracted with
the metric $\eta^{A_w B_w}$). The  bosonic  term is manifestly invariant under the
triality exchange of the three groups in $SL_0(2,R) \times SL_1(2,R) \times SL_w(2,R)$: it means that 
any permutation of the {\small $A_0,A_1,A_w$} indices leaves 
the action invariant. 

It is easy to verify that the action is invariant under the supersymmetry transformations (\ref{SBA}).

Notice however, that the bosonic  term
and the fermionic term are separately invariant under the reshuffling of the $SL(2)$ indices. For the action to be invariant under the same triality, we have to identify the groups $SL_i(2,R)$, $i=1,2,3$. Namely, the action is invariant only under the small triality reshuffling and not under the
big pentality reshuffling of the fermionic terms.  To see this,
consider the bosonic coordinates $X_{A_0 A_1 A_w}$.
The action is invariant for example under the reshuffling $X_{A_0 A_1 A_w} \rightarrow X_{A_1 A_w A_0}$ as discussed above. However, if we are reshuffling the indices as  $X_{A_0 A_1 A_w}
\rightarrow X_{A_0 A_2 A_w}$, where we exchange $SL_1(2)$ with $SL_2(2)$, we have to define
the new quantities  $X_{A_0 A_2 A_w}$ since they are now charged under a new $SL(2)$.
So, in order to complete the triality, we have to identify $X_{A_0 A_2 A_w}$
with $X_{A_0 A_1 A_w}$, which means that they transform only under the diagonal subgroup of
$SL_1(2) \times SL_2(2)$. Adding also $X_{A_0 A_3 A_w}$ we obtain an action invariant under
$SL_0 \times SL_{diag} \times SL_w$, $SL_{diag}$ being the diagonal subgroup of $SL_1(2) \times SL_2(2) \times SL_3(2)$.
 In the same way we proceed for the fermions. 

We can therefore rewrite the action as
\begin{eqnarray}\label{SBB}
S_m 
&=& \frac{1}{3}  \int d^2x\,  
\Big[ 
\eta^{A_w B_w} \e^{A_0 B_0} 
(\e^{A_1 B_1} X_{A_0 A_1 A_w} X_{B_0 B_1 B_w} + 
\e^{A_2 B_2} X_{A_0 A_2 A_w} X_{B_0 B_2 B_w} \nonumber \\
&+&
\e^{A_3 B_3} X_{A_0 A_3 A_w} X_{B_0 B_3 B_w} ) +
 \eta^{A_w B_w} \e^{A_0 B_0} 
(\e^{A_2 B_2}  \e^{A_3 B_3}
\Theta_{A_0 A_2 A_3 A_w} \Theta_{B_0 B_2 B_3 B_w}  
\nonumber \\
&+&
\e^{A_1 B_1}  \e^{A_2 B_2}
\Theta_{A_0 A_1 A_2 A_w} \Theta_{B_0 B_1 B_2 B_w}  +
\e^{A_1 B_1}  \e^{A_3 B_3}
\Theta_{A_0 A_1 A_3 A_w} \Theta_{B_0 B_1 B_3 B_w})
\Big]\,. \nonumber
\end{eqnarray}

The triality under the exchange of $SL_0$, $SL_{diag}$ and $SL_w$ becomes manifest 
after adding some boundary terms, as we did in the case of the NSR action. This means adding to the metric the $B$ term as in
(\ref{LCG}),  replacing  the invariant tensors $\eta^{A_w B_w} \e^{A_0 B_0} 
\e^{A_i B_i}$ with $\frac{1}{3}( 
\eta^{A_w B_w} \e^{A_0 B_0} \e^{A_i B_i} + 
\eta^{A_0 B_0} \e^{A_i B_i} 
\e^{A_w B_w} + 
\eta^{A_i B_i} \e^{A_w B_w} 
\e^{A_0 B_0}  
)
$ , 
$ \eta^{A_w B_w} \e^{A_0 B_0} 
\e^{A_2 B_2}  \e^{A_3 B_3}
$ with $\frac{1}{3} 
(\eta^{A_w B_w} \e^{A_0 B_0} \e^{A_2 B_2} + 
\eta^{A_0 B_0} \e^{A_2 B_2} 
\e^{A_w B_w} + 
\eta^{A_2 B_2} \e^{A_w B_w} 
\e^{A_0 B_0})  $ $\times  \e^{A_3 B_3}
$ 
and similarly for the terms $ \eta^{A_w B_w} \e^{A_0 B_0} 
\e^{A_1 B_1}  \e^{A_3 B_3}$ and $ \eta^{A_w B_w} \e^{A_0 B_0} 
\e^{A_1 B_1}  \e^{A_2 B_2}
$.
 In this way we add only boundary terms. 
 
For the ghost field, the 
situation is more involved, but since this is a consequence of the specific gauge choice that reduces
the Green-Schwarz action to the Siegel-Berkovits action, the possible violation of the triality is only through BRST 
exact terms which do not affect the physical amplitudes. 

There are two important aspects that we have to point out. The first one is that the boundary terms
for the fermionic pieces work as in the case of the bosonic terms, and therefore the action is manifestly
invariant under the triality that exchanges the three groups $SL_0(2)$, $SL_w(2)$ and $SL_{diag}(2)$.
The second aspect, as was noted in \cite{Berkovits}, is that
the choice ${\cal N} = 2+2$ is mandatory to cancel the BRST anomaly.
Here we found that the triality -- which is only present in the case of the supergroup $OSp(2,2|2)$ --
implies that cancellation of the anomalies. This is a confirmation of  previous work and
an unexpected present from the triality. There is an additional minor point: the fermionic terms display an additional $SL(2,R)$ symmetry which implies a tetrality instead of a triality. We do not have any interpretation, but it might refer to a twist between the R-symmetry and
the triality.

 The present formulation is suitable for computations of
amplitudes and the manifest duality should show up in the computations. This will be explored in a separate work.

\sect{Super-Hyper-Det based on $SL(2|1)^3$ algebra}

Given the results of the previous sections it is natural to try
and generalize the hyperdeterminant, invariant under $SL(2)^3$, to
a supersymmetric object, invariant under a superalgebra which
contains $SL(2)^3$ as a bosonic subalgebra. In fact we can build a
quartic (bosonic) supersymmetric object based on the $SL(2|1)^3$
superalgebra which, by setting a suitable set of fields to zero,
precisely reproduces the hyperdeterminant of \cite{Cayley}.\\ The
$SL(2|1)^3$ superalgebra is made out of three copies of the
following:
\begin{align}
\left\lbrace Q_A, Q_B \right\rbrace &= P_{AB} \nn \\ \left[
P_{AB}, Q_C \right] &= -\epsilon_{C(A} Q_{B)} \\ \left[ P_{AB},
P_{CD} \right] &= 2 \epsilon_{(A(C} P_{D)B)} \nn
\end{align}
The indices {\small $A,A',A''$} label the three $SL(2|1)$ factors of the
superalgebra. It is possible to construct a $SL(2|1)^3$
representation with 27 fields, 14 of which, $X_{AA'A''}$, $Y_A$,
$Y_{A'}$ and $Y_{A''}$, are bosonic while the remaining 13,
$\psi_{AA'}$, $\psi_{A'A''}$, $\psi_{A''A}$ and $\eta$ are
fermionic. The action of the algebra on the fields is given by:
\begin{align*}
Q_{A} X_{BB'B''} &= \frac{1}{2} \epsilon_{AB} \psi_{B'B''} &
Q_{A'} X_{BB'B''} &= \frac{1}{2} \epsilon_{A'B'} \psi_{B''B} &
Q_{A''} X_{BB'B''} &= \frac{1}{2} \epsilon_{A''B''} \psi_{BB'} \\
Q_{A} \psi_{BB'} &= \epsilon_{AB} Y_{B'} & Q_{A'} \psi_{BB'} &= -
\epsilon_{A'B'} Y_{B} & Q_{A''} \psi_{BB'} &= X_{BB'A''}\\ Q_{A}
\psi_{B'B''} &= X_{AB'B''} & Q_{A'} \psi_{B'B''} &=
\epsilon_{A'B'} Y_{B''} & Q_{A''} \psi_{B'B''} &= -
\epsilon_{A''B''} Y_{B'}\\ Q_{A} \psi_{B''B} &= - \epsilon_{AB}
Y_{B''} & Q_{A'} \psi_{B''B} &= X_{BA'B''} & Q_{A''} \psi_{B''B}
&= \epsilon_{A''B''} Y_{B}\\ Q_{A} Y_{B} &= \frac{1}{2}
\epsilon_{AB} \eta & Q_{A'} Y_{B} &= - \frac{1}{2} \psi_{BA'} &
Q_{A''} Y_{B} &= \frac{1}{2} \psi_{A''B} \\ Q_{A} Y_{B'} &=
\frac{1}{2} \psi_{AB'} & Q_{A'} Y_{B'} &= \frac{1}{2}
\epsilon_{A'B'} \eta & Q_{A''} Y_{B'} &= - \frac{1}{2}
\psi_{B'A''} \\ Q_{A} Y_{B''} &= - \frac{1}{2} \psi_{B''A} &
Q_{A'} Y_{B''} &= \frac{1}{2} \psi_{A'B''} & Q_{A''} Y_{B''} &=
\frac{1}{2} \epsilon_{A''B'} \eta \\ Q_{A} \eta &= Y_{A} & Q_{A'}
\eta &= Y_{A'} & Q_{A''} \eta &= Y_{A''}
\end{align*}
In the quartic invariant, only the following bilinear building
blocks contribute:
\begin{align*}
X_{(AB)} &= X_{AA'A''} X_{BB'B''} \epsilon^{A'B'}
\epsilon^{A''B''} & A_{(AB)} &= \psi_{AA'} \epsilon^{A'B'}
\psi_{BB'} \\ B_{(AB)} &= \psi_{A''A} \epsilon^{A''B''}
\psi_{B''B} & W_{(AB)} &= Y_A Y_B \\ \omega_A &= Y_{A''}
\epsilon^{A''B''} \psi_{B''A} & \nu_A &= Y_{A'} \epsilon^{A'B'}
\psi_{AB'} \\ \Delta_A &= X_{AA'A''} \psi_{B'B''} \epsilon^{A'B'}
\epsilon^{A''B''} & \chi_{A} &= Y_A \eta
\end{align*}
together with their prime and double prime counterparts; notice
that the building blocks with two indices are bosonic and those  with one index are
fermionic. These blocks can be rearranged in the
combinations: $$\mathcal{Z}_{AB} \equiv 2 X_{AB} - A_{AB} - B_{AB}
- 2 W_{AB}$$ $$\Phi_A \equiv 2 \sx( \D_A - \n_A + \o_A - \chi_A
\dx)$$ which obey very simple supersymmetry relations: $$Q_A
\mathcal{Z}_{BC} = \e_{A(B} \Phi_{C)}$$ $$Q_A \Phi_B =
\mathcal{Z}_{AB}$$ $$Q_{A'} \mathcal{Z}_{AB} = Q_{A''}
\mathcal{Z}_{AB} = 0$$ $$Q_{A'} \Phi_A = Q_{A''} \Phi_A = 0 $$
Then one easily checks that
\begin{equation}
H = -\frac{1}{48} \sx( \mathcal{Z}_{AB} \mathcal{Z}^{AB} +
\mathcal{Z}_{A'B'} \mathcal{Z}^{A'B'} + \mathcal{Z}_{A''B''}
\mathcal{Z}^{A''B''}  + \Phi_A \Phi^A +
\Phi_{A'} \Phi^{A'} + \Phi_{A''} \Phi^{A''} \dx)
\nonumber \end{equation}
\noi is invariant under the action of the superalgebra. The
indices are raised/lowered with the use of the $SL(2)$--invariant
epsilon tensors according to the rule given in (\ref{SLmetric}),
and the factor $-\frac{1}{48}$ has been chosen to reproduce the
hyperdeterminant once all the fields but $X_{AA'A''}$ are set to
zero.
 \sk
 \newpage
  {\bf Note 1:} $H$ can be seen as the definition of the
  super-Cayley determinant of
  the cubic supermatrix given in Fig. 1:
\sk
\let\picnaturalsize=N
\def\picsize{2.5in}
\def\picfilename{cubo1.eps}
\ifx\nopictures Y\else{\ifx\epsfloaded Y\else\input epsf \fi
\let\epsfloaded=Y
\centerline{\ifx\picnaturalsize N\epsfxsize \picsize\fi
\epsfbox{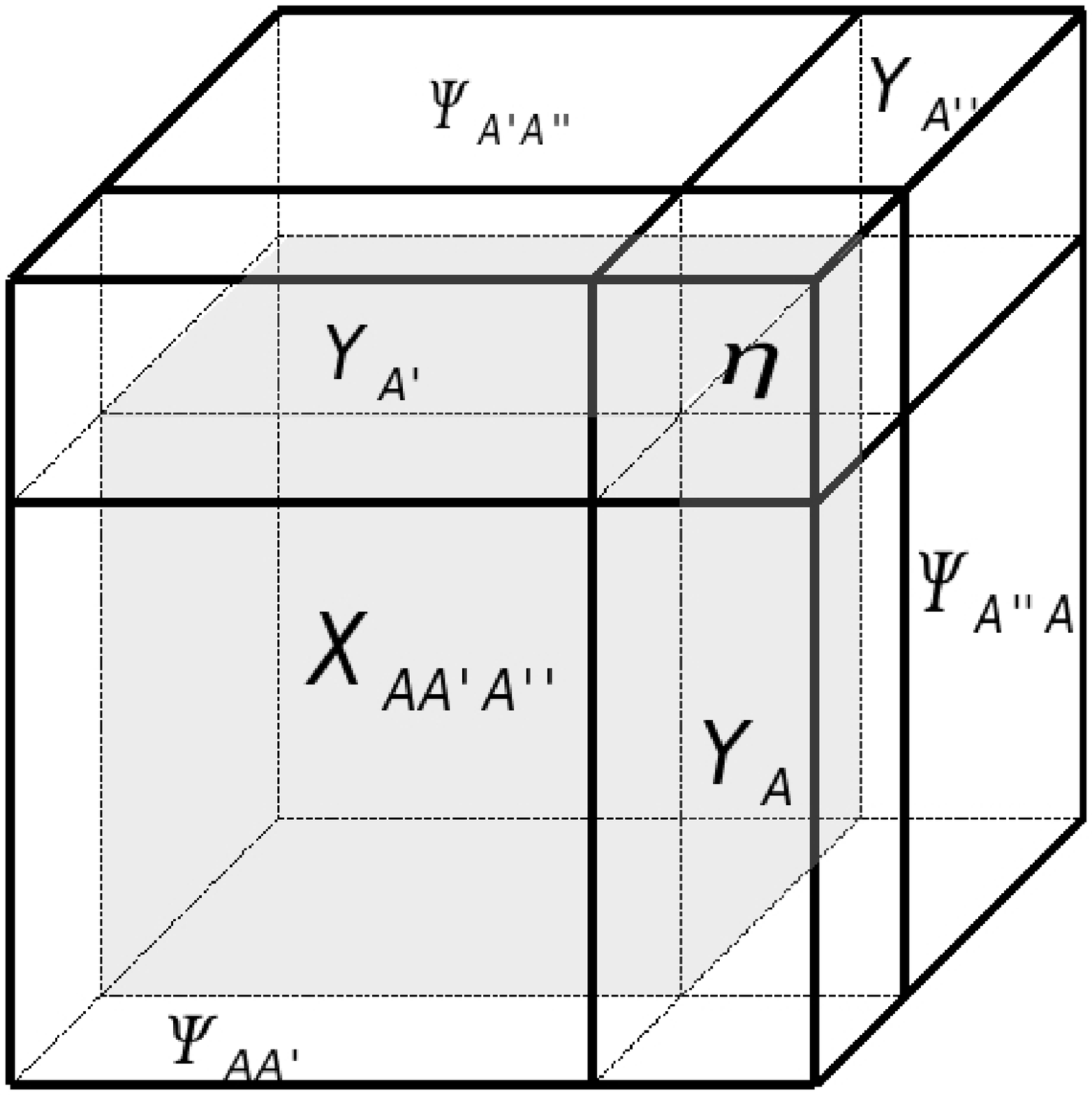}}}\fi
  {\small{\bf Fig. 1} : the $3 \times 3 \times 3$ cubic supermatrix}

 \sk
  {\bf Note 2:} $H$ is also equal to the sum of the Berezinians of the three $3 \times 3$ supermatrices

\begin{equation}
\begin{pmatrix}
\mathcal{Z}_{AB} & {1 \over \sqrt{2}}  \Phi_A \\   {1 \over \sqrt{2}} \chi_B & 1
\end{pmatrix}
\qquad
\begin{pmatrix}
\mathcal{Z}_{A'B'} &  {1 \over \sqrt{2}} \Phi_{A' }\\   {1 \over \sqrt{2}} \chi_{B' }& 1
\end{pmatrix}
\qquad
\begin{pmatrix}
\mathcal{Z}_{A''B''} &  {1 \over \sqrt{2}} \Phi_{A''}\\ {1 \over \sqrt{2}} \chi_{B''} & 1
\end{pmatrix}
\end{equation}

\noi with $\chi_B \equiv - \Phi^C \mathcal{Z}^{-1}_{BC}$ etc.

\sect{Conclusions and outlook}

We have constructed the supersymmetric generalization of the triality invariance first found by Duff in the Nambu-Goto string moving 
in a flat $D=2+2$ target space. This we achieve by adding boundary terms in the NSR superstring action, and in the Siegel-Berkovits formulation of the selfdual superstring. Moreover we have
proposed a supersymmetric generalization of the Cayley hyperdeterminant, based on a quartic invariant of the
$SL(2|1)^3$ superalgebra. It may be intriguing to speculate on its possible applications
in quantum information or in the description of black holes in string/brane theory.

\appendix

\section{$D=2$ gamma matrices}
We use the representation:
\begin{equation}
\g^0 =
\begin{pmatrix}
0 & - \ii \\
\ii & 0
\end{pmatrix}
\qquad
\g^1 =
\begin{pmatrix}
0 & \ii \\
\ii & 0
\end{pmatrix}
\end{equation}
for the two--dimensional $\g$--matrices, satisfying the usual
relations $$\sx\{ \g^\a, \, \g^\b \dx\} = - \eta^{\a\b} \quad
{\mathrm{and}} \quad \g^\a \g^\b = - \eta^{\a\b} \iden + \e^{\a\b}
\g_3$$ where the metric is $\eta = \sx( - , + \dx)$, $\e$ is the
usual Levi--Civita symbol and $$\g_3 =  \begin{pmatrix} 1 & 0
\\ 0 & -1\end{pmatrix}$$ The charge conjugation matrix is $C =
\g^0$, so that all the spinors are real and the following
relations hold: $$\g^0 \sx( \g^\a \dx)^\dagger \g^0 = \g^\a,
\qquad \g^0 \sx( \g^\a \dx)^t \g^0 = - \g^\a$$ Finally, for
Majorana fermions the currents satisfy:
\begin{align}
\ol \xi \zeta &= \ol \zeta \xi \\
\ol \xi \g_3 \zeta &= - \ol \zeta \g_3 \xi \\
\ol \xi \g^\a \zeta &= - \ol \zeta \g^\a \xi
\end{align}
The $SL(2)$-invariant tensor $\e^{\a\b}$ is used to raise and
lower the indices according to:
\begin{equation}
V_\a = \e_{\a\b} V^\b \qquad V^\a = - \e^{\a\b} V_\b \label{SLmetric}
\end{equation}

\sect{Some notes on $OSp(2,2|2)$}

The supergroup is characterized by the following superalgebra generated  by the bosonic generators $P_{AB},P'_{A'B'},P''_{A''B''}$ and by the fermionic  generators $Q_{AA'A''}$:
$$ \Big\{ Q_{AA'A''}, Q_{B B' B''} \Big\} =
{1 \over 2} \e_{AB} \e_{A'B'} P_{A'' B''} +  {1 \over 2}\e_{AB} P'_{A'B'} \e_{A'' B''} -P_{AB} \e_{A'B'} \e_{A'' B''} \,, $$
\begin{equation}
[P_{AB}, P_{CD} ] =2 \e_{(A (C} P_{D) B)}
\label{saA}
\end{equation}
$$[P'_{A'B'}, P'_{C'D'} ] = 2 \e_{(A' (C'} P'_{D') B')}\,, $$
$$[P''_{A''B''}, P''_{C''D''} ] = 2 \e_{(A'' (C''} P''_{D'') B'')}\,, $$
$$[P_{AB}, Q_{C C' C''}] = - \e_{C (A}  Q_{B) C' C''}\,, $$
$$[P'_{A'B'}, Q_{C C' C''}] = - \e_{C' (A'}  Q_{C | B') C''}\,, $$
$$[P''_{A''B''}, Q_{C C' C''}] = - \e_{C'' (A''}  Q_{C C' | B'')}\,, $$

They provide the adjoint representation of the superalgebra. Denoting by $T_{\cal M}$ the supergenerators of $OSp(2,2|2)$, by $V_{\cal M}$ the components of the supermultiplet and by $f_{\cal M N}{}^{\cal R}$ the super-structure constants, we set
\begin{equation}
\label{saAB}
T_{\cal M} V_{\cal N} = f_{\cal M N}{}^{\cal R} V_{\cal R}\,.
\end{equation}
and it is obvious to see that it forms a representation. Notice that since the representation is linear, there is no problem to set either $X_{AA'A''}$ as a fermion or as a boson.

\end{document}